\documentclass[a4paper,fleqn,usenatbib]{mnras}

\usepackage{newtxtext,newtxmath}

\usepackage[T1]{fontenc}
\usepackage{ae,aecompl}

\usepackage{graphicx}	
\usepackage{amsmath}	
\usepackage{amssymb}	

\title[Phaethon surface inhomogeneity]{Rotational variation of the linear polarisation of the 
asteroid (3200) Phaethon as evidence for inhomogeneity in its surface properties\thanks{Based on data collected with 2-m RCC telescope at Rozhen National Astronomical Observatory and Omicron (west) telescope of the C2PU facility at the Calern observing station of the Observatoire de la C\^ote d'Azur and GOTO Telescope operated on the island of La Palma}}

\author[G. Borisov et al.]{
G. Borisov,$^{1,2}\thanks{Galin.Borisov@Armagh.ac.uk (GB)}$
M. Devog\`{e}le,$^{3}$
A. Cellino,$^{4}$
S. Bagnulo,$^{1}$
A. Christou,$^{1}$
Ph. Bendjoya,$^{5}$
\newauthor
J.-P. Rivet,$^{5}$
L. Abe,$^{5}$
D. Vernet,$^{5}$
Z. Donchev,$^{2}$
Yu. Krugly,$^{6}$
I. Belskaya,$^6$
T. Bonev,$^2$
\newauthor
D. Steeghs,$^{7}$ 
D. Galloway,$^{8}$ 
V. Dhillon,$^{9}$ 
P. O'Brien$^{10}$,
D. Pollacco,$^{7}$ 
S. Poshyachinda,$^{11}$ 
\newauthor
G. Ramsay,$^{1}$
E. Thrane,$^{8}$ 
K. Ackley,$^{8}$ 
E. Rol,$^{8}$
K. Ulaczyk,$^{7}$ 
R. Cutter$^{7}$, 
M. Dyer$^{9}$
\\
$^{1}$Armagh Observatory and Planetarium, College Hill, Armagh BT61 9DG, Northern Ireland, United Kingdom\\
$^{2}$Institute of Astronomy and NAO, Bulgarian Academy of Sciences, 72, Tsarigradsko Chauss\'ee Blvd., BG-1784 Sofia, Bulgaria\\
$^{3}$Lowell Observatory, 1400 West Mars Hill Road, Flagstaff, AZ 86001, U.S.A. \\
$^{4}$INAF - Osservatorio Astrofisico di Torino, via Osservatorio 20, 10025 Pino Torinese (TO), Italy\\
$^{5}$Universit\'e C{\^o}te d'Azur, Observatoire de la C{\^o}te d'Azur, France\\
$^{6}$Institute of Astronomy of Kharkiv National University, Sumska Str. 35, Kharkiv 61022, Ukraine\\
$^{7}$Astronomy and Astrophysics Group, Department of Physics, University of Warwick, Gibbet Hill Road, Coventry, CV4 7AL, United Kingdom\\
$^{8}$Monash Centre for Astrophysics, School of Physics \& Astronomy, Monash University, Clayton, VIC 3800, Australia\\
$^{9}$Department of Physics and Astronomy, University of Sheffield, Sheffield, S3 7RH, United Kingdom\\
$^{10}$Department of Physics and Astronomy, University of Leicester, Leicester LE1 7RH, United Kingdom\\
$^{11}$National Astronomical Research Institute of Thailand, 191 Siriphanich Bldg, Huay Kaew Road, Muang District, Chiang Mai 50200, Thailand
}
\date{Accepted XXX. Received YYY; in original form ZZZ}

\pubyear{2018}

\begin{document}
\label{firstpage}
\pagerange{\pageref{firstpage}--\pageref{lastpage}}
\maketitle

\begin{abstract}
Asteroid (3200) Phaethon is a Near-Earth Apollo asteroid with an unusual orbit that brings it closer 
to the Sun than any other known asteroid. Its last close approach to the Earth was in mid-December 2017 
and the next one will be on October 2026. Previous rotationally time-resolved spectroscopy 
of Phaethon showed that its spectral slope is slightly bluish, in agreement with its B/F taxonomic 
classification, but at some rotational phases, it changes to slightly reddish. Motivated by this 
result we performed time-resolved imaging polarimetry 
of Phaethon during its recent close 
approach to the Earth. Phaethon has a spin period of 3.604\,hours and we found a variation of 
the linear polarisation with rotation. 
This seems to be a rare case in which such variation is unambiguously found, also a consequence 
of its fairly large amplitude. 
Combining this new information with the brightness and colour variation, as well as previously 
reported results from Arecibo radar observations, we conclude that there is no variation of the
mineralogy across the surface of Phaeton. However, the observed change in the linear polarisation 
may be related to differences in the thickness of the surface regolith in different areas or local 
topographic features.
\end{abstract}




\section{Introduction}
Linear polarisation of sunlight scattered by asteroid surfaces is routinely measured and shows 
changes with the phase angle (the Sun--Target--Observer angle).
While it is well known that the intensity of scattered sunlight varies with viewing geometry and 
object shape, only in a few cases a periodic variation of the 
degree of linear polarisation, synchronous with the spin period 
of the object, is observed. 
The asteroid Vesta is such a case and it has long been known to be the only asteroid exhibiting 
such a variation, with the same period as the object's rotation 
\citep[see][and references therein]{CellinoVesta}. This variation must be the consequence of some 
heterogeneity of the asteroid's surface, for example the existence of regions characterised by different 
albedo, or mineralogy, 
or regolith properties, or a combination of the above. Some authors have reported the possible detection 
of a weak polarimetric modulation for a few other objects, including for example (9) Metis, (52) Europa, 
and (1036) Ganymed \citep{Nakayama2000}, but Vesta is the only asteroid for which this phenomenon 
has been convincingly demonstrated, confirmed at different epochs and interpreted in terms of local surface 
properties directly determined by {\em in situ} measurements by the Dawn space probe \citep{CellinoVesta}.

Asteroid (3200) Phaethon is the parent body of the Geminid meteor shower 
\citep{Whipple, Fox, Green, Gustaf, WW}, and the real physical 
nature of this object (asteroid or comet?) has been a long-debated subject \citep{comet}. 
It is also an Apollo asteroid 
(Near-Earth-asteroid (NEA) with semi-major axis larger than Earth's $a>1$\,au, 
but perihelion distances $q<1.017$\,au) 
with an unusual orbit that brings it closer to the Sun than any other named asteroid ($q\simeq0.14$\,au). Its last 
close approach to the Earth was in December 2017 (16 Dec 2017 23:00, $\Delta$=0.0689\,au) 
and the next one is not until October 2026. Recently, \citet{Kinoshita2017} reported rotationally 
time-resolved spectroscopy of Phaethon. Their main result is that the slope of the asteroid's spectrum 
is slightly blue, in agreement with its known B/F-type taxonomy, 
but at some rotational phases, it changes to slightly red. 
Those authors' interpretation is that Phaethon has a red spot on its surface. 
Also, \citet{3200Nature} published polarimetric observations of Phaethon during its last apparition. 

Motivated by \citet{Kinoshita2017} developments we decided to perform time-resolved imaging polarimetry and 
photometry in different filters and with three different instruments. We want to to investigate if there 
is a surface inhomogeneity and understand, in particular, which effects, including possible 
variations in composition and/or in size or shape of the surface 
regolith particles, can be responsible of what is observed.

\section{Observations and Data reduction}
\subsection{Observations}
A multi-colour phase-polarisation curve of Phaethon has been obtained during 
the December 2017 apparition by merging measurements taken with the 2-channel focal reducer 
(FoReRo2) \citep{FoReRo2} attached to the 2-m Ritchey-Chr\'etien-Coud\'e (2mRCC) telescope 
at the Bulgarian National Astronomical Observatory (BNAO) - Rozhen and with the Torino Polarimeter 
(ToPol) \citep{ToPol} mounted on the Omicron (west) telescope of the C2PU facility at the Calern 
observing station of the Observatoire de la C\^ote d'Azur. 

All the observations were obtained in the 
positive polarisation branch, where the electric field vector of reflected light from the asteroid 
preferably oscillates perpendicularly to the scattering plane, with the phase angle ranging from 36$^\circ$ 
to 116$^\circ$. For this investigation we use a subset 
of this data obtained on the night of 15 December 2017, when we observed the object continuously 
to cover a full rotational period of 3.604\,hours 
\citep[Light Curve Database (Rev. 2018-March);][]{LCDB,PhaethonShape}. 
The observations were obtained simultaneously using polarimetric instruments in Rozhen and 
Calern observatories in Johnson-Cousins $BVRI$ filters. The log of our polarimetric observations 
is given in Table~\ref{Tab:data}.

In order to compare variability of the polarisation with the asteroid's rotation, we carried out photometry 
simultaneously with the polarimetry observations to obtain a lightcurve of the asteroid 
and to see if there is a variation of its colour with rotation as well. For this purpose we 
used the Calern telescope with its wide-field camera and the Gravitational-wave Optical Transient Observer  
(GOTO) at Roque de Los Muchachos observatory on La Palma.
The photometric observations from Calern were obtained in SDSS $g'$ and $r'$ filters a few days earlier, 
on 12 December 2017. 

GOTO is a multi-telescope facility on La Palma in the Canary Islands whose prime 
aim is to detect optical counterpart of gravitational wave events. At the time of the observations being 
reported here, three 0.4\,m telescopes were attached to a common mount, each giving a field of view 
of 2.1$\times$2.8\,deg and pixel size 1.2$^{''}$/pixel. It is therefore well suited to study fast moving 
sources like Phaeton. Using broadband Baader $LRGB$ filters during dark conditions 
we typically reach $\sim$20.5 magnitude in the $L$ band with 5\,min exposure time. An overview of GOTO 
and a detailed description of the telescope control and scheduling system can be found in \citet{GOTO}. 
The GOTO observations were made on 15 December 2017 using $RGB$ filters.

\begin{table}
\caption{The log of the photometric and polarimetric observations of (3200) Phaethon 
obtained at Rozhen, Calern and GOTO observatories.}
\centering                          
\begin{tabular}{cccc}
\hline \hline
Date & Obs. Type & Filter & Observatory \\
\hline
 12 Dec 2017 & photometry & SDSS $g'r'$ & Calern \\
 15 Dec 2017 & photometry & Baader $RGB$ & GOTO \\
 15 Dec 2017 & polarimetry & $R$ & Rozhen \\
 15 Dec 2017 & polarimetry & $BVRI$ & Calern \\
\hline
\end{tabular}
\label{Tab:data}
\end{table} 

\subsection{Data reduction}
Standard reduction steps, including bias subtraction and flat-fielding were performed for all the images 
obtained by all the instruments.

\subsubsection{Photometry}
Photometry was done using aperture photometry with a circular aperture with size 
2$\times$FWHM. 
Photometry calibration was done using field stars cross-matched with the APASS \citep{APASS} and 
the Pan-STARRS \citep{PanSTARRS1,PanSTARRS2} catalogues, for GOTO and Calern data respectively. 
The average atmospheric extinction for both observing sites was used to compute the above-atmosphere 
instrumental magnitudes. In order to convert the GOTO magnitudes from the 
$RGB$ passbands to SDSS $g'$ and $r'$, colour-colour and magnitude-magnitude relations were 
constructed and fitted using 84 cross-matched stars in the field. 

Baader $B$ and $G$ filters are relatively narrow (B: 3850-5100\,\AA, G: 4950-5750\,\AA) 
and both of them together cover only the wavelength range of SDSS $g'$. 
In addition Baader $R$ filter (R: 5850-6900\,\AA) overlaps with SDSS $r'$. 
Consequently, we can compute the $(g'-r')$ colour 
in 3 different ways and the SDSS $g'$ magnitude in 2 different ways using Baader $B$ and $G$. 

The following equations were used for the calibration:
\begin{eqnarray}
\begin{array}{ccrcr}
(g'-r')&=&T_{BR}(B-R)_0&+&C_{BR} \\
(g'-r')&=&T_{GR}(G-R)_0&+&C_{GR} \\
(g'-r')&=&T_{BG}(B-G)_0&+&C_{BG} \\
r'-R_0&=&T_R(g'-r')&+&ZP_{Rr'} \\
g'-G_0&=&T_G(g'-r')&+&ZP_{Gg'} \\
g'-B_0&=&T_B(g'-r')&+&ZP_{Bg'} 
\end{array}
\label{GOTOeq}
\end{eqnarray}
and the fitted values of the coefficients are listed in Table~\ref{GOTOtab}.
\begin{table}
	\centering
	\caption{The coefficients of the transformation from GOTO $BGR$ magnitudes to SDSS $g'$ and $r'$ 
	ones -- Zero points and slope parameters as defined in Equation~\ref{GOTOeq}.}
	\begin{tabular}{rcrccrcr} 
		\hline
$T_{BR}$ &=& 0.769$\pm$0.016 &&& $C_{BR}$ &=& 0.313$\pm$0.006 \\
$T_{GR}$ &=& 1.344$\pm$0.031 &&& $C_{GR}$ &=& 0.398$\pm$0.005 \\
$T_{BG}$ &=& 1.327$\pm$0.030 &&& $C_{BG}$ &=& 0.267$\pm$0.007 \\
$T_R$ &=& 0.028$\pm$0.022 &&& $ZP_{Rr'}$ &=& 21.469$\pm$0.012 \\
$T_G$ &=& 0.451$\pm$0.021 &&& $ZP_{Gg'}$ &=& 21.673$\pm$0.012 \\
$T_B$ &=& -0.173$\pm$0.026 &&& $ZP_{Bg'}$ &=& 21.824$\pm$0.014 \\
		\hline
	\end{tabular}
\label{GOTOtab}
\end{table}

\subsubsection{Polarimetry}
The flux of the object at each polarised beam was calculated using a growth-curve method but 
instead of following the growth of the total intensity with aperture size, we follow the stokes parameters 
$X/I = (S_{\parallel}-S_{\perp})/(S_{\parallel}+S_{\perp})$. This method usually gives the saturation at 
smaller aperture, which means that we introduce less noise. The polarisation then was computed 
applying the so called "beam-swapping technique" \citep{Bagnulo}. The results were 
corrected for the instrumental polarisation using unpolarised standard stars observed during the same night.

\begin{figure}
\includegraphics[angle=0,width=0.45\columnwidth]{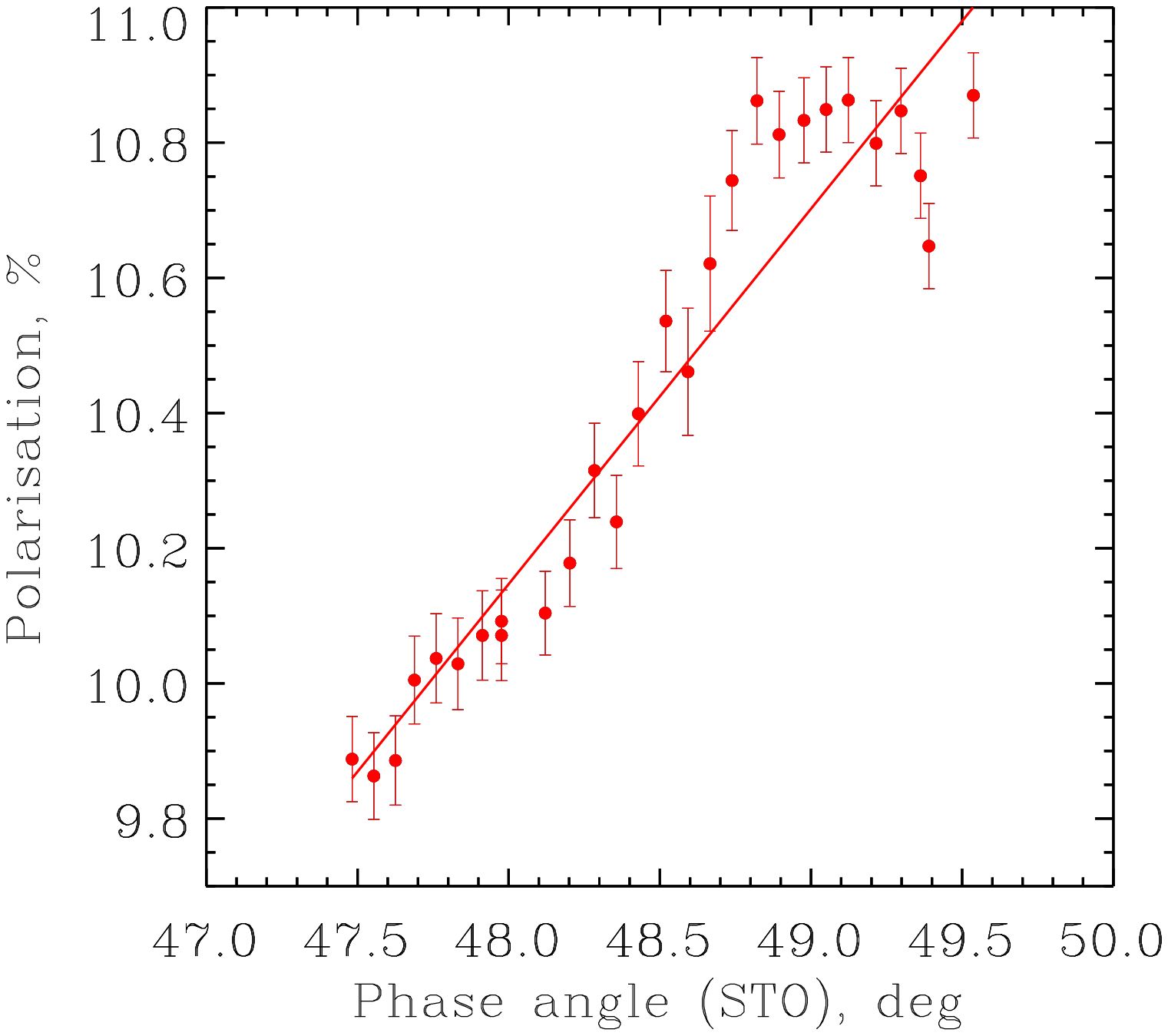}
\includegraphics[angle=0,width=0.45\columnwidth]{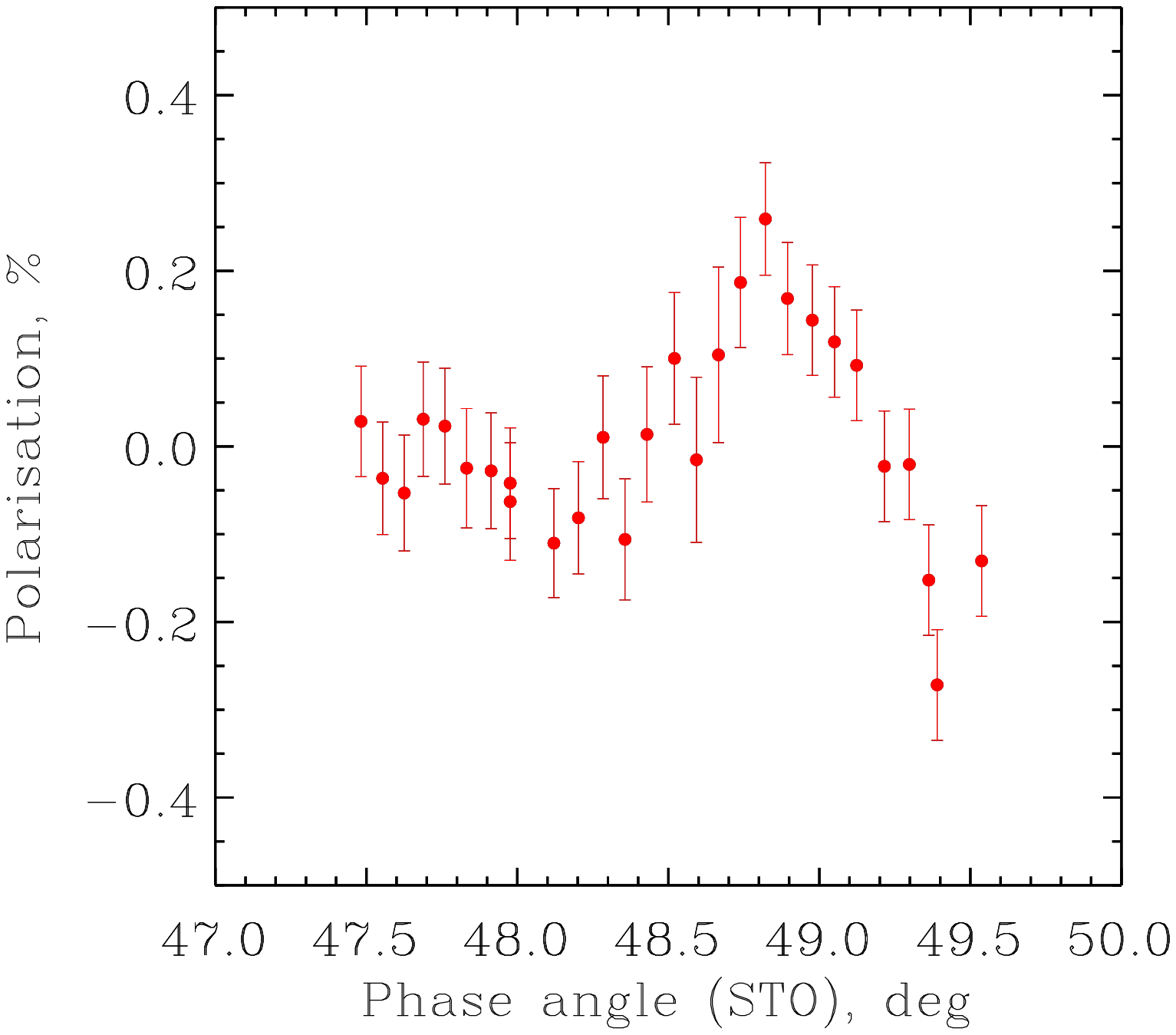}
\caption{The phase angle (Sun--Target--Observer) dependence on the  polarisation is presented 
in the left panel of the figure. The points represent the SDSS $r'$ filter measurement from Rozhen 
with 1-$\sigma$ error bars. 
The solid line is the linear least square fit to the data, which is the behaviour we expect 
for this range of phase angles. The right panel represents the polarisation residuals after 
subtracting the best fit curve from the data.}
\label{PolPhase}
\end{figure}

As the asteroid passed by the Earth at a very small distance, the phase angle changed significantly 
during the observations, so the dependence of the polarisation and brightness on it during the 
night was stronger than on the rotation phase (see left panel of Fig.~\ref{PolPhase}). 
The values of linear polarisation we obtained are among the highest ever observed 
for a low-albedo near-Earth asteroid. The interval of phase angle was not sufficiently covered to 
derive a firm determination of the $P_{max}$ parameter, but this appears to occur at a phase angle 
around 130$^\circ$. For more details see \citet{PhaethonPol}. 
At the values of phase angle used for this investigation, the 
linear polarisation is expected to show a smooth, linear variation with increasing phase, 
but our measured values are not distributed as smooth as expected. 
A modulation synchronous with the asteroid rotation is superimposed on the linear change of linear polarisation. 
Therefore, in order to isolate the polarisation-rotation variability a linear least square fit was made to 
the polarisation-phase angle variation and removed from the data. The residuals between our data and fit 
is presented in the right panel of Fig.~\ref{PolPhase}. Then the rotational phase 
for each epoch of observations was calculated and the rotational variation of the polarisation 
was constructed (see Fig.~\ref{polrot}). 

\section{Results}
\begin{figure*}
\includegraphics[angle=90,width=0.84\textwidth]{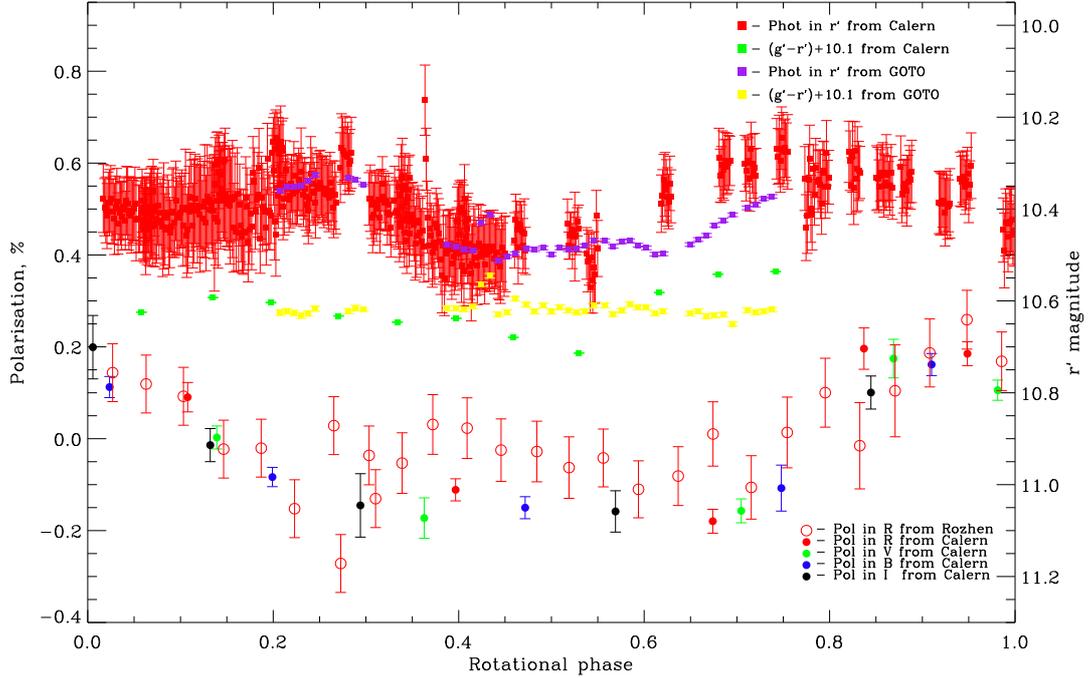}
\caption{Comparison of the (3200) Phaethon SDSS $r'$ lightcurve (from Calern in red and 
GOTO in purple) and $(g'-r')$ (in green for Calern and in yellow for GOTO) and the variation of 
the degree of linear polarisation obtained from Rozhen and Calern in different filters (see text and 
labels for details). All error bars represent 1-$\sigma$ uncertainties.}
\label{polrot}
\end{figure*}

Figure~\ref{polrot} presents the results of our observations of Phaethon. The lightcurves obtained 
from C2PU and GOTO are presented with red and purple squares 
respectively and are mutually consistent within the 1-$\sigma$ uncertainties. The slight difference 
between C2PU and GOTO for rotational phases above 0.6 most probably is caused by rapidly changing 
geometry for these observations. The aspect angle (the angle between the asteroid spin vector 
and the line of sight to the observer) 
is changing by 10 degrees, computed using the axis orientation from the latest shape model by \citet{PhaethonShape}. 
The measured colour of the object, obtained with the same instruments, apparently shows no variation 
with rotation. On the other hand, the residual linear polarisation, after removing the phase angle 
dependence trend, changes with the asteroid's rotation. It is presented in Fig.~\ref{polrot} with open 
red circles for Rozhen and colour-coded filled circles for each different filter for Calern. The observed 
trend of polarisation is clearly not due to measurement noise as the amplitude of the variability is 
$>$5 times of the measurement uncertainties. It is mostly anticorrelated with the photometric 
lightcurve (see Fig.~\ref{polrot}), the maximum of polarisation being reached at the minimum of the 
lightcurve, and vice versa, except for the interval of rotation phase between about 0.65 and 0.80, 
where the correlation is positive.

\section{Discussion}
Information about asteroid shapes can be obtained by classical photometric observations, under 
the assumption that the measured brightness variation is due to a periodically 
changing cross-section. One single lightcurve can provide a  determination 
of the spin period, and some crude indication about the overall shape (a large lightcurve 
amplitude being a consequence of an elongated shape seen in favourable observing conditions). 
An accurate and reliable determination of the shape usually requires 
lightcurves obtained at different apparitions, corresponding to different aspect angles. Provided 
a sufficient number of lightcurves is available, this also allows to determine the orientation of the 
spin axis. However, it is generally difficult to identify from photometric lightcurves alone definitive 
evidence of a possible surface albedo heterogeneity due to 
variations in surface composition and/or regolith properties. This is the reason why polarimetry 
is so important. The polarisation state of the scattered surface, in fact, strongly depends 
on surface albedo and regolith structure \citep{Cellino2016}. 

The anticorrelation between brightness and polarisation of Phaethon 
visible in Fig.~\ref{polrot} is most simply interpreted by Umov's law \citep{Umov}, 
which states that the degree of polarisation is proportional to the reciprocal value of the albedo, 
in other words {\it high polarisation$\rightarrow$low albedo$\rightarrow$low brightness}. 
For example, \citet{CellinoVesta}, comparing linear polarisation measurements of 
Vesta with albedo maps from {\it Dawn} space mission, showed that the maximum of 
linear polarisation occurs when the regions of lowest albedo are facing 
the observer. 
As a consequence, if the same phenomenon was at work in the case of Phaethon, we would be 
led to conclude that we find evidence of albedo variations across the surface of this asteroid. 
On the other hand, the fact that we didn't see any variation of the colour of the object with rotation 
suggests that there might be no variation of the mineralogy as a function of longitude. 
\citet{3200Nature} discuss that high polarisation, for a fixed albedo, should be caused by 
large particles (d$\sim$360\,$\mu$m). This is also supported by 
the blue slope of its spectrum, which usually is interpreted also in terms of larger grain sizes \citep{B-Slope}. 

A possible solution to this apparent discrepancy may be found by considering another result 
by \citet{CellinoVesta} for Vesta, namely that 
the thickness of surface regolith can play a role in determining the fraction of linear 
polarisation of the scattered sunlight. In particular, these authors suggest that high elevation areas on Vesta, 
expected to have a thinner regolith due to a higher gradient of topographic slope favouring 
the accumulation 
of regolith particles at lower altitudes, are associated with higher linear polarisation. 
This effect seems to be so strong as to compete with the anti-correlation between 
polarisation and albedo, leading to local violation of Umov's law \citep{CellinoVesta}. 
This phenomenon, qualitatively detected so far only for Vesta, seems to be at work also for  
Phaethon, if we look at the positive correlation between polarisation 
and brightness at rotational phases between 0.65 and 0.80 (see Fig.~\ref{polrot}). In contrast 
to Vesta in the case of Phaethon we have no information 
about local topography and we cannot make a generic interpretation, 
in terms of the existence of a local region on Phaethon's surface characterised by 
different regolith properties. 
Recent Arecibo radar observations of Phaethon show a concavity, or depression, 
at least several hundred meters in extent, near the equator and a radar dark feature near one of the 
poles \citep{Arecibo}. Those features might be responsible for polarisation variation caused by the 
processes explained above.

\section{Conclusions}

There is a clear variation of the degree of linear polarisation with rotational phase of asteroid 
(3200) Phaethon and it is mostly anticorrelated with its brightness variation except in 
a small interval of rotation phase. On the other hand there is no significant variation of the $(g'-r')$ colour with the rotation.

This anticorrelation can be explained by Umov's law.
We found similarities with polarisation behaviour of asteroid Vesta, which leads us to interpret our 
observations in a similar way i.e. some form of heterogeneity of the asteroid's surface, such as the 
existence of regions characterised by different albedo, different regolith properties, or both. 
A less likely interpretation, not supported by our colour measurements, is a variation of the mineralogy. 
We also observe that Umov's law is violated for a narrow range of rotational phase and, similar to 
the case of Vesta, can be explained by either a small region with a steep topographic slope on the surface 
with a very thin regolith layer or a spot with different surface composition. 
Arecibo radar observations of Phaethon show a dark spot on it surface and a concavity region near 
the equator. 

The results of our analysis of (3200) Phaethon, together with the results recently published by  
\citet{3200Nature} from its previous apparition and previous results for Vesta \citep{CellinoVesta} 
suggest that polarimetry is diagnostic not only of surface albedo, but also of other properties 
of the surface regolith. This is not, strictly speaking, a new discovery, since it was already known 
that some polarimetric properties 
(primarily the $P_{min}$ -- $\alpha_{inv}$ relation, where $P_{min}$ is the extreme 
value of polarisation in the negative branch of the phase-polarisation curve and $\alpha_{inv}$ 
is the inversion phase angle where the polarisation is zero and the phase-polarisation curve goes 
from negative to positive branch and vice versa, see \citet{Cellino2016} for a recent reassessment of the subject) 
are diagnostic of properties such as the regolith particle size. 
We can conclude therefore that we are at the beginning 
of a new era in asteroid polarimetry. Effects that were previously 
underestimated are now more routinely found to play an important 
role in asteroid polarimetric measurements. This is largely 
a result of the availability of new and better 
detectors and of a renewed interest in this field of investigation 
by several research teams.

\section*{Acknowledgements}

This work was supported via a grant (ST/M000834/1) from the UK Science and Technology Facilities 
Council. 
We gratefully acknowledge observing grant support from the Institute of Astronomy and Rozhen 
National Astronomical Observatory, Bulgarian Academy of Sciences. 
The Calern Asteroid Polarimetric Survey (CAPS), carried out at Calern in the framework of C2PU, 
is a collaboration between INAF - Torino Astrophysical Observatory and the Observatoire de la cote d'Azur. 
The GOTO Observatory is a collaboration between the University of Warwick and Monash University 
(as the Monash-Warwick Alliance), Armagh Observatory \& Planetarium, the University of Sheffield, 
the University of Leicester, the National Astronomical Research Institute of Thailand (NARIT), 
the Instituto de Astrofsica de Canarias (IAC), the University of Turku and Rene Breton (University of Manchester). 
TB acknowledges financial support by contract DN 18/13-12.12.2017 with the Bulgarian NSF.
This research was made possible through the use of the AAVSO Photometric All-Sky Survey (APASS), 
funded by the Robert Martin Ayers Sciences Fund. 
The Pan-STARRS1 Surveys (PS1) and the PS1 public science archive have been made possible through contributions by the Institute for Astronomy, the University of Hawaii, the Pan-STARRS Project Office, the Max-Planck Society and its participating institutes, the Max Planck Institute for Astronomy, Heidelberg and the Max Planck Institute for Extraterrestrial Physics, Garching, The Johns Hopkins University, Durham University, the University of Edinburgh, the Queen's University Belfast, the Harvard-Smithsonian Center for Astrophysics, the Las Cumbres Observatory Global Telescope Network Incorporated, the National Central University of Taiwan, the Space Telescope Science Institute, the National Aeronautics and Space Administration under Grant No. NNX08AR22G issued through the Planetary Science Division of the NASA Science Mission Directorate, the National Science Foundation Grant No. AST-1238877, the University of Maryland, Eotvos Lorand University (ELTE), the Los Alamos National Laboratory, and the Gordon and Betty Moore Foundation.
\vspace{-0.6cm}
\bibliographystyle{mnras}
\bibliography{Phaethon} 
%
%
\bsp	
\label{lastpage}
\end{document}